\documentclass[%
 reprint,
superscriptaddress,
 amsmath,amssymb,
 aps,
]{revtex4-2}
\usepackage{xcolor}
\usepackage{graphicx}
\usepackage{dcolumn}
\usepackage{bm}

\usepackage[caption = false]{subfig}
\usepackage{float}
\begin{document}

\preprint{APS/123-QED}

\title{Dynamics of elongation of nematic tactoids in an electric field }

\author{Mohammadamin Safdari}
\author{Roya Zandi}%
\affiliation{%
Department of Physics, University of California, Riverside, California 92521, USA
}%
\author{Paul van der Schoot}
\affiliation{
Department of Applied Physics and Science Education, Eindhoven University of Technology, Eindhoven, The Netherlands\\
}%


\date{\today}

\begin{abstract}
Nematic tactoids are spindle-shaped droplets of a nematic phase nucleated in the co-existing isotropic phase. According to equilibrium theory, their internal structure and shape are controlled by a balance between the elastic deformation of the director field, induced by the preferred anchoring of that director field to the interface, and the interfacial free energy. Recent experiments on tactoids of chitin nanocrystals dispersed in water show that electrical fields can very strongly elongate tactoids, at least if the tactoids are sufficiently large in volume. However, this observation contradicts the predictions of equilibrium theory as well as findings from Monte Carlo simulations that do not show this kind of extreme elongation to take place at all. To explain this, we put forward a relaxational model based on the Oseen-Frank free energy of elastic deformation of a director field coupled to an anisotropic surface free energy. 
In our model, we use two reaction coordinates to describe the director field and the extent of elongation of the droplets, and evaluate the evolution of both as a function of time following the switching on of an electric field. Depending on the relative magnitude of the fundamental relaxation rates associated with the two reaction coordinates, we find that the aspect ratio of the drops may develop a large and very long-lived overshoot before eventually relaxing to the much smaller equilibrium value.  In that case, the response of the curvature of the director field lags behind, explaining the experimental observations. Our theory describes the experimental data reasonably well.
\end{abstract}

\maketitle

\section{\label{sec:level1}Introduction}
Dispersions containing elongated colloidal particles have long been known to form nematic liquid-crystalline phases if their volume or weight fractions exceeds some critical value that actually can be surprisingly small~\cite{dogic2016filamentous}. As the transition from the isotropic to a nematic state is discontinuous, the coexistence of these two states occurs in a range of concentrations in between the binodals associated with them~\cite{van2022molecular}. The development of a  macroscopic nematic phase under conditions of coexistence with the isotropic phase is rather slow, however, and involves the formation of nematic droplets called tactoids in the isotropic parent phase~\cite{Jamali2015,van2006isotropic,van1996liquid,oakes2007growth,almohammadi2023disentangling}. Nematic tactoids have an unusual, elongated shape reminiscent of a spindle, a rugby ball or an American football, reflecting the underlying uniaxial symmetry of the nematic phase. That the structure and properties of tactoids may be studied experimentally is because it may take months to years before the nematic droplets sediment and coalesce, and a single nematic phase presents itself. The nematic phase is typically denser than the co-existing isotropic phase~\cite{van2022molecular}.

Since the pioneering work of Z\"{o}cher on vanadium pentoxide sols in 1925~\cite{Zocher1925}, tactoids have been observed in a wide range of molecular, polymeric and colloidal lyotropic liquid-crystalline fluids. These include dispersions of different kinds of filamentous and rod-like virus particles~\citep{BAWDEN1936,dogic2003,modlinska2015,tarafder2020phage}, inorganic nanorods~\cite{Coper1937,sonin2017mineral}, polypeptides~\cite{Robinson1956}, carbon nanotubes~\cite{puech2010,Jamali2015}, F-actin~\cite{oakes2007}, protein amyloids \cite{bagnani2018amyloid}, chromonic liquid crystals~\cite{Kim2013} and cellulose nanocrystals~\cite{REVOL1992,kitzerow1994,park2014,wang2016}. The characteristic shapes and director-field structures of nematic tacoids have been the subject of a large number of theoretical and computer simulation studies~\cite{Kaznacheev2002,Prinsen2003,Prinsen2004a,Prinsen2004b,kaznacheev2003,Lettinga2005,Lettinga2006,Williams1986,van2012,otten2012,Everts2016,almohammadi2022shape}. From these studies we understand that the spindle shape so typical of tactoids must be due to the competition between the preference of rod-like particles for a planar anchoring of the director field to the interface, and the resulting elastic deformation of the (quasi bipolar) director field. Because the free energy associated with the elastic deformation of the director field and the interfacial free energy depend differently on the volume, both the shape and director field configuration depend on the volume of a tactoid~\cite{Prinsen2004a}. 

Experimental findings on the structure and shape of tactoids seem to be reasonably well described by macroscopic theory, even if they are not very large on the scale of the length of the colloidal particles. Indeed, the application of macroscopic theory makes possible the extraction of information on the elastic constants of the nematic and interfacial free energies between the co-existing isotropic and nematic phases from polarization microscopic images alone if the experimental data include a sufficiently wide range of tactoid sizes~\cite{Jamali2015,kaznacheev2003,Prinsen2003,Prinsen2004a,Prinsen2004b,safdari2021,Metselaar2017,weirich2017liquid}.
Interestingly, 
the experiments on the aspect ratio (or length and breadth) as a function of the volume of tactoids typically exhibit a significant amount of scattered data. Puech \textit{et al.}~\cite{puech2010} argued that this must be due to thermal fluctuations on account of the fact that the interfacial free energies of lyotropic nematics must be very low. It so happens that for this type of colloidal system interfacial tensions can be of the order of $\mu$N m$^{-1}$ or even significantly below that~\cite{puech2010,van1999,chen2002interfacial,van2006isotropic}. Incidentally, this also explains why tactoids require very little energy to significantly deform, and why the shape and structure of tactoids are so strongly affected by the contact with an adsorbing surface or the presence of an externally applied electric, magnetic or flow field~\cite{almohammadi2020flow,Jamali2017,Metselaar2017,kaznacheev2003}. 

\begin{figure}[!h]
\includegraphics[width=7.2cm]{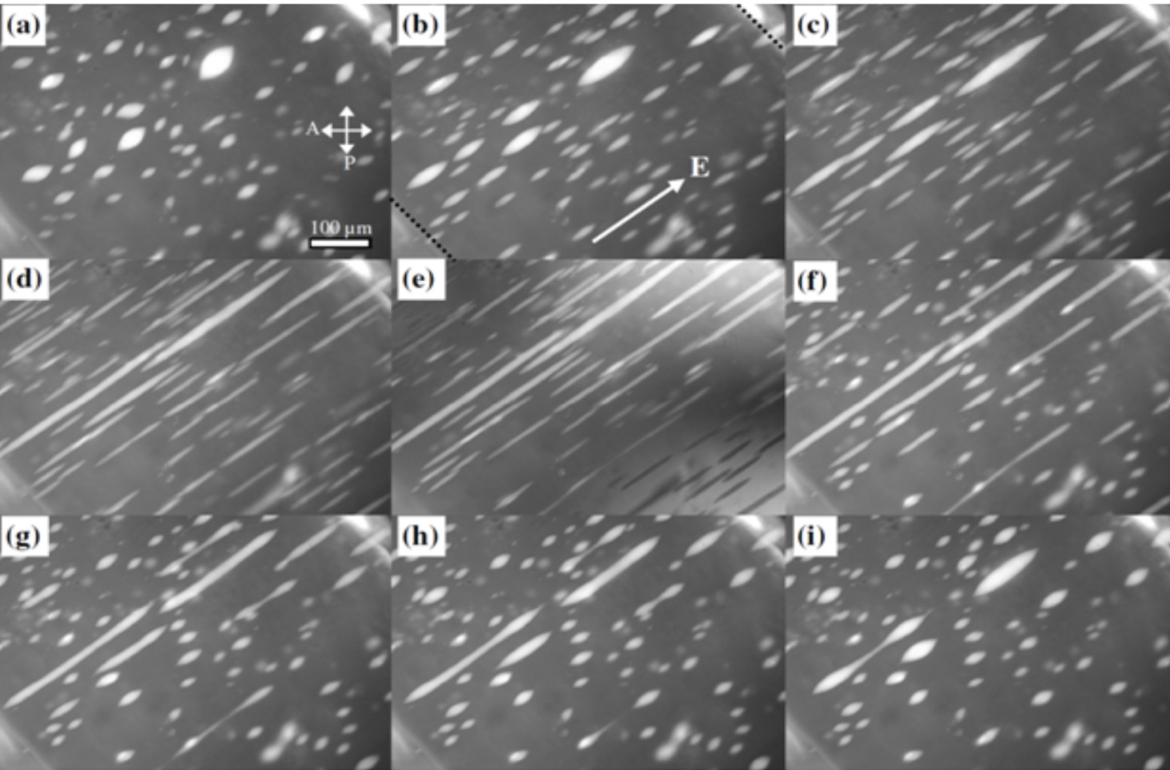}\caption{Polarization microscopic images of the evolution in time of tactoids in an aqueous solution of chitin fibers, following the switching on and off of an AC electric field of root-mean-square magnitude of $E=160$ V mm$^{-1}$ and a frequency of $300$ kHz. (a) to (e): The field is applied at time zero, and a collection of tactoids imaged at 0, 120, 210, 640, and 1080 seconds. (f) to (i): The field is switched off and the same tactoids are imaged after 230, 380, 530, and 820 seconds following the removal of the electric field. Figure reproduced with permission from Ref.~\cite{Metselaar2017}.}	 \label{Experimental_of_Metselaar}
\end{figure}

It is also clear from the work of Jamali {\it et al}.~on nematic tactoids found in dispersions of carbon nanotubes in a superacid, however, that this variation {\it cannot} be explained in terms of the magnitude of thermal (equilibrium) fluctuations predicted by macroscopic theory~\cite{Jamali2015}. This already suggests that the equilibration of the shape and director field of tactoids might be quite slow~\cite{almohammadi2022shape}. In fact, recent work by Mezzenga and collaborators on tactoids in aqueous dispersions of amyloids and of cellulose nanocrystals supports this: tactoids produced via nucleation and growth may have a different aspect ratio and internal structure than those produced in a microfluidic device~\cite{almohammadi2023disentangling}. This explains to some extent why experiments on tactoids in aqueous dispersions of chitin in the presence of an applied AC electric field~\cite{Metselaar2017} show a large disconnect with predictions from equilibrium theory and the results from Monte Carlo simulations~\cite{safdari2021,kuhnhold2022structure}. Experimentally, tactoids are able to elongate by a factor of up to about ten through their coupling to the electric field. In contrast, in equilibrium theory and Monte Carlo simulations it is the director field that responds to the external orienting field with the aspect ratio of the tactoids showing much less sensitivity, see also Figure~\ref{Experimental_of_Metselaar}.

In Refs.~\cite{safdari2021,kuhnhold2022structure} this discrepancy was elucidated by suggesting that the response of droplet shape and the director field occurs on different time scales: Initially the anchoring enslaves the director field and forces it to follow the droplet shape. Only after the droplet shape has relaxed, then the director field starts relaxing. This eventually leads to the final relaxation of the droplet shape. which in turn leads to the subsequent relaxation of the shape of the drop to the actual equilibrium value in the very much later stages of the process. Indeed, Mezzenga and collaborators recently also suggested that tactoid equilibration might be kinetically controlled, leading to very long-lived metastable states~\cite{bagnani2018amyloid}. This, then, would also explain the large scatter in tactoid shape and director field in dispersions that have been left to equilibrate for a long time, in that full equilibrium might not yet have been reached for the tactoids themselves. This is remarkable, given that they typically measure from ten to a few hundreds of $\mu$m in length~\cite{Jamali2015,Jamali2017}. 

In this article, we follow up on our previous work~\cite{safdari2021}, where we investigated the equilibrium shape and structure of nematic tactoids in an external alignment field, and introduce a relaxational dynamics based on two reaction coordinates. These reaction coordinates describe (i) the elongation of a tactoid with a prescribed spindle shape, and (ii) the degree of deformation of the director field where we prescribe the geometry of the deformation. The prescribed droplet shape and director-field geometry allows us to straightforwardly evaluate the well-known Oseen-Frank elastic free energy as well as the surface free energies that we use as input for our dynamical theory. For the surface free energy, we use the ansatz of Rapini and Papoular~\cite{rapini1969}. By construction, the steady-state solution to our equations produces the optimal aspect ratio and degree of curvature of the director field, given the volume of the drop and the strength of the alignment field~\cite{safdari2021}. 

We find that the elongation of tactoids in an electric field is \textit{entirely} a kinetic effect. Depending on the ratio of the two fundamental relaxation rates associated with the two reaction coordinates, we either obtain a monotonic time evolution of both the aspect ratio and curvature of the director field, or a monotonic response of the curvature of the director field and an overshoot in the aspect ratio of the tactoids. The actual response times depend strongly on the strength of the electric field and the volume of the tactoids, appropriately scaled to the elastic constants and interfacial free energies associated with the nematic. Curve-fitting to the experimental data of Metselaars and collaborators~\cite{Metselaar2017} gives reasonable agreement, showing that the relaxation of the bipolarness of the director field must be extremely sluggish. The cause of this remains unclear and requires further study.

The remainder of this paper is structured as follows. In Section II we describe the ingredients of our theory. Section III summarizes our main findings based on a numerical evaluation of the theory. In Section IV we compare our theory with experiments, and finally we present our conclusions and discuss our results in the Section V.

\section{Theory}
Relaxational or model A dynamics is based on a phenomenological description of how a system relaxes to a state of equilibrium, described by an appropriate free energy\cite{Hohenberg, clark2023relaxational,kra2023energetics}. The free energy of the nematic droplet is the sum of three components, 
\begin{equation}
F = F_{S} + F_{E} + F_{C},
\end{equation}
where $F_{S}$, $F_{E}$ and $F_{C}$ refer to the contributions of (i) the interface between the nematic tactoid and the host isotropic phase it is in contact with, (ii) the elastic deformation of the director field in the tactoid and (iii) the interaction of the nematic with alignment field, respectively. 

\begin{figure}
    \centering
    \includegraphics[width=8cm]{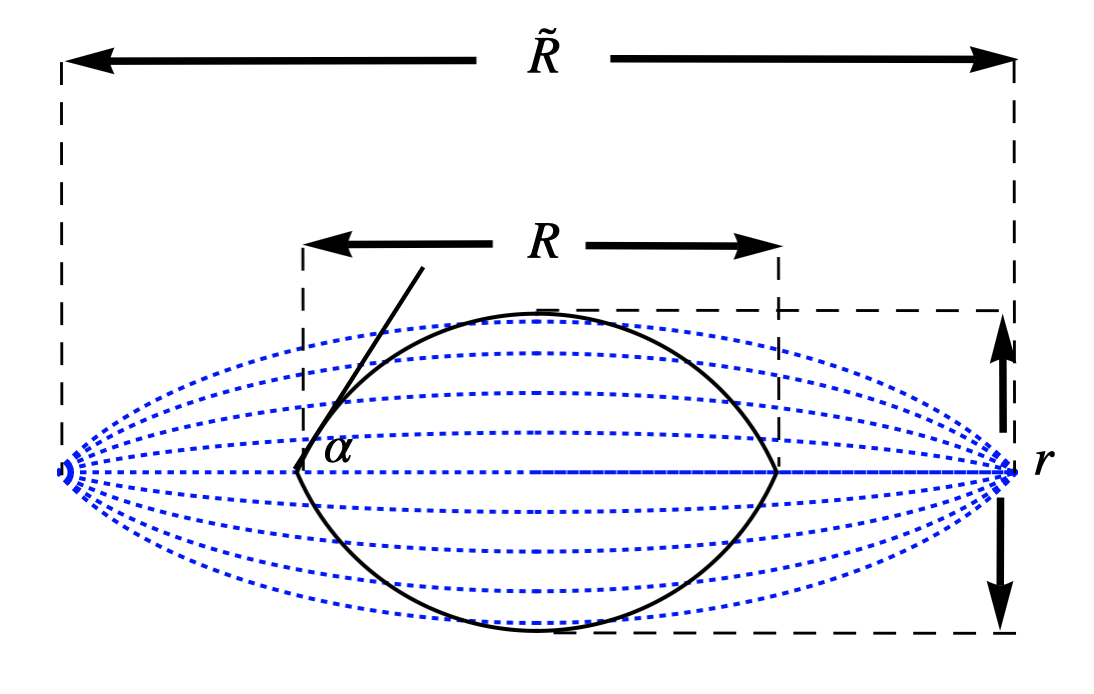}
    \caption{Cross section (solid) and director field (dotted) of the structure of a tactoid presumed in our calculations. The droplet is cylindrically symmetric about its main axis. $R$ denotes the length of a tactoid and $r$ its width. $\tilde{R}$ is the distance between the virtual boojums, which are the focal points of the (extrapolated) bipolar director field. For $R=\tilde{R}$, the virtual boojums become actual boojums, {\it i.e.}, surface point defects. 
    Also indicated is the opening angle $\alpha$ of the spindle-shaped droplet. See also the main text.
    }
    \label{fig0}
\end{figure}

For the interfacial free energy, we use the well-known phenomenological expression of Rapini and Papoular~\cite{free,rapini1969}, 
\begin{equation}
\begin{aligned}
F_S =&\sigma \int \left[1+\omega(\vec{q} \cdot \vec{n})^2 \right] \mathrm{d}A, 
\end{aligned}
\label{FrankFreeEnergy}
\end{equation}
that in the context of dispersions of rod-like particles has some merit~\cite{van1999}. Here, $\sigma$ denotes the interfacial tension between the nematic droplet and isotropic medium for the case of perfect planar anchoring, $\vec{q}$ is the local surface normal, $\vec{n}$ the director field at the surface $A$ and $\omega>0$ a dimensionless anchoring strength that penalizes non-planar anchoring of the director field to the  interface. The integration is over the entire surface $A$ of the droplet. For simplicity, we presume the interfacial tension and anchoring strength to be independent of the strength of the alignment field, even though the isotropic phase becomes paranematic in such a field and we would expect the interfacial tension and anchoring strength to respond to that~\cite{finner2020geometric}. As usual, we also ignore any curvature dependence of the interfacial free energy and invoke the so-called capillarity approximation.

For the elastic deformation, we make use of the Oseen-Frank free energy~\cite{frank1958liquid},
\begin{equation}
\begin{aligned}
F_E = &\int    \left(\frac{1}{2} K_{11} \left( \vec{\nabla} \cdot \vec{n} \right)^2 + \frac{1}{2}K_{33} \left( \vec{n} \times \left( \vec{\nabla} \times \vec{n} \right) \right)^2 
\right.
\\
& - \left.
K_{24}\vec{\nabla} \cdot \left(\vec{n} \vec{\nabla} \cdot \vec{n} + \vec{n} \times \left( \vec{\nabla} \times \vec{n} \right) \right)\right) \mathrm{d} V ,
\end{aligned}
\label{FrankFreeEnergy}
\end{equation}
where $K_{11}$, $K_{33}$, and $K_{24}$ represent the elastic modulus of splay, bend, and saddle-splay terms respectively. In our description, we do not allow for a twisted director field~\cite{Prinsen2004b}. Integration is over the entire volume $V$ of the drop. We take the elastic moduli to be independent of the strength of the field. Strictly speaking this cannot be true as their values depend on the degree of nematic alignment of the particles~\cite{odijk1986}. For the kind of bipolar director field that we shall be considering, the saddle-splay deformation renormalizes the splay elastic contribution to the free energy, implying that $K_{11}$ becomes $K_{11}-K_{24}$ and we can remove the saddle-splay term from Eq.~\ref{FrankFreeEnergy}~\cite{Prinsen2003,Prinsen2004a}.

In the experiments of Metselaar and collaborators~\cite{Metselaar2017}, the alignment field is a high-frequency AC electric field that we treat as a static field $\vec{E}$ with a magnitude equal to the root-means-square value of that of the AC field. At time zero we switch on the field, and are interested in the response of  tactoids to the switching on of the field. We expect the dielectric susceptibility of the nematic phase to be anisotropic and reflect its uniaxial symmetry~\cite{de1993}. Hence, the dielectric susceptibility of the nematic can be described as a second-rank tensor with two principal susceptibilities $\epsilon_{\parallel}$ parallel and $\epsilon_{\perp}$ and perpendicular to the local director. Defining the susceptibility anisotropy as $\Delta \epsilon\equiv \epsilon_{\parallel}-\epsilon_{\perp} \geq 0 $, the electric-field (or Coulomb) contribution to the free energy becomes 
\begin{equation}
F_{C}= - \dfrac{1}{8 \pi} \Delta \epsilon \int \left(\vec{n} \cdot \vec{E} \right)^2 \mathrm{d}V,
\label{FreeEnergyElectricField}
\end{equation}
apart from a constant term that we can ignore~\cite{safdari2021,toropova2022nucleation}. The integration is again over the entire volume $V$ of the tactoid. The anisotropy of the dielectric susceptibility depends on the degree of nematic ordering and hence also depends, in principle, on the strength of the electric field. This, we also ignore.

Obviously, we do not know the shape and director-field configuration \textit{a priori}. As in our previous work, we follow Prinsen \textit{et al.}~and Kaznacheev \textit{et al.}~and choose to prescribe the shape of the droplet and the geometry of the director field~\cite{Prinsen2004a,kaznacheev2003,safdari2021}. For the shape we use a circle section of revolution, and for the director field a bipolar director field, sketched in Fig.~\ref{fig0}. The shape and director field are then fully described by the length of the drop $R$, the width of the drop $r$, and the distance between the focal points of the bipolar director field $\tilde{R}$. The aspect ratio of a tactoid, $x\equiv R/r$, is related to the opening angle $\alpha$, also indicated in Fig.~\ref{fig0}, via the relation $x=\cot{\alpha/2}$. The degree of bipolarness of the director field is described by the quantity $y = \tilde{R}/R$. For a homogeneous director field $y \rightarrow \infty$ whilst for a purely bipolar field $y \rightarrow 1$. In the latter case the foci of the bipolar director field are located on the poles of the tactoid and represent surface point defects called boojums~\cite{volovik1983topological}. We refer to our earlier work~\cite{safdari2021} and also to Fig.~\ref{fig0}. Note that for any value of the bipolarness $1<y<\infty$, the director field is \textit{quasi} bipolar.

It turns out that the free energy can be entirely described in terms of the variables $\alpha = 2 \cot^{-1} x$ and $y$, and in terms of four dimensionless groups describing all the elastic and surface materials parameters of the model. The volume of a droplet $V$ is fixed during an experiment and also constitutes a model parameter that we make dimensionless by defining $v\equiv \left(\sigma/(K_{11}-K_{24})\right)^3V$, recalling that $\sigma$ denotes the bare surface tension and $K_{11}$ and $K_{24}$ are the elastic constants associated with a splay and saddle-splay deformation of the director field. 
The free energy we also render dimensionless. A definition that proves to be practical is $f\equiv \sigma F/\left(K_{11}-K_{24}\right)^2$. 

With these definitions, the full expression of our free energy now becomes~\cite{safdari2021}
\begin{equation}
\begin{aligned}
f(\alpha,y)=&\quad   \mathrm{v}^{2/3}\phi^{-2/3}_\mathrm{v}(\alpha)(\phi_\sigma(\alpha)+\omega \phi_{\omega} (\alpha,y))
\\
&+\mathrm{v}^{1/3}\phi_\mathrm{v}^{-1/3}(\alpha)(\phi_{11}(\alpha,y) + \kappa \phi_{33}(\alpha,y))
\\
&
-\Sigma \mathrm{v}\phi_\mathrm{v}^{-1}(\alpha)\phi_{C}(\alpha,y) 
\label{free-energy-dimensionless}
\end{aligned}
\end{equation}
The first term of Eq.~\ref{free-energy-dimensionless} represents the contribution from the surface tension and anchoring energy, with
\begin{equation}
\phi_\mathrm{v}(\alpha) = \frac{7\pi}{3}+ \frac{\pi}{2}\left( \frac{1 - 4\alpha \cot \alpha + 3\cos 2\alpha}{\sin^2 \alpha}\right),
\end{equation} 
\begin{equation}
\begin{aligned}
\phi_\sigma(\alpha) = 4\pi \left(  \frac{1-\alpha \cot\alpha}{\sin \alpha}\right),
\end{aligned}
\end{equation}
and
\begin{equation}
\begin{aligned}
\phi_{\omega}(\alpha,y) & = \frac{\pi}{2} (y^2-1)^2 \sin^3 \alpha \\ & \times \int_{0}^{\pi} \mathrm{d} \xi \left[  \dfrac{\sin^2 \xi  \cos^2 \xi   }{N(y,\xi,\alpha)\left(1+\sin \xi \cos \alpha \right)^3} \right].
\end{aligned}
\label{eqanchoring}
\end{equation}
Here,
\begin{equation}
\begin{aligned}
N(y,\xi,\eta)  =  & \left(\sin \xi  \cos \eta + \frac{1}{2} Z(\xi,\eta) \left( y^2 - 1 \right) \right)^2 \\ & + y^2 \sin^2 \xi \sin^2 \eta,
\end{aligned}
\label{eqN}
\end{equation}
and
\begin{equation}
Z(\xi,\eta) =  1 + \sin \xi \cos \eta,
\end{equation}
where we set $\eta = \alpha$ in Eq.~\ref{eqN}. 

The second term of Eq.~\ref{free-energy-dimensionless} is due to the Frank elastic energy, where the dimensionless group $\kappa = K_{33}/(K_{11}-K_{24})$ acts as a measure for the importance of the bend elastic deformation.
Here,
\begin{equation}
\begin{aligned}
\phi_{11}(\alpha,y) =& 8\pi  \int_{0}^{\pi} \mathrm{d} \xi \int_{0}^{\alpha} \mathrm{d} \eta \sin^2 \xi  \cos^2 \xi  \sin \eta \\
&\times \frac{1}{N(y,\xi,\eta)\left(1+\sin \xi \cos \eta \right)^3},
\end{aligned}
\end{equation}
describes the contribution of the splay and saddle-splay deformations, and
\begin{equation}
\begin{aligned}
\phi_{33}(\alpha,y) =& 8\pi \int_{0}^{\pi} \mathrm{d} \xi \int_{0}^{\alpha} \mathrm{d} \eta   \sin^4 \xi  \sin^3 \eta \\ & \times \frac{1}{N(y,\xi,\eta)\left(1+\sin \xi \cos \eta \right)^3},
\end{aligned}
\end{equation}
that of the bend deformation. 

Finally, the third term describes the contribution from the interaction of the nematic droplet with an electric field, with $\Sigma \equiv \frac{1}{8\pi} \epsilon_\mathrm{a} E^2 \sigma^{-2}(K_{11}-K_{24})$ a measure for its strength relative to the surface and elastic deformation free energy cost, and
\begin{equation}
\begin{aligned}
 \phi_{\mathrm{C}}(\alpha,y)   = 8\pi \int_{0}^{\pi} \mathrm{d} \xi \int_{0}^{\alpha} \mathrm{d} \eta  \frac{\sin^2 \xi  \sin \eta }{(1+\sin \xi \cos \eta )^3} \\
 \times \frac{\left(y^2 Z^2+\sin^2 \xi \sin^2 \eta -\cos^2 \xi \right)^2}{N(y,\xi,\eta)\left(1+\sin \xi \cos \eta \right)^2} .
\end{aligned}
\end{equation}

We have not been able to analytically evaluate the various integrals, except when the tactoids are extremely elongated and the opening angle becomes very small. In that case, asymptotic relations may be obtained as well as robust scaling estimates~\cite{Prinsen2004a,safdari2021}. Since we do not wish to restrict ourselves to large aspect ratios, we rely on a numerical evaluation of the integrals. For this, we employ the Mathematica software package~\cite{wolfram1999mathematica} and use the $NIntegrate$ function to compute integrals and the $D$ function to calculate derivatives. 

Now that we have formulated our free energy in terms of the two reaction coordinates $\alpha$ and $y$, we are able to formulate our relaxational theory in terms of the generalized forces:
\begin{equation}
\begin{aligned}
& \dfrac{\partial \alpha}{\partial t} = - \tilde{\Gamma}_{\alpha} \dfrac{\partial F}{\partial \alpha},
\\
&\dfrac{\partial y}{\partial t} = - \tilde{\Gamma}_{y} \dfrac{\partial F}{\partial y},
\end{aligned}
\label{kineticequations}
\end{equation}
where we choose to ignore the contribution of  cross terms~\cite{Doi2011Onsager}, the main reason being to limit the number of adjustable parameters in our model. Here, $\tilde{\Gamma}_{\alpha}$ and $\tilde{\Gamma}_{y}$ are fundamental relaxation rates that have dimensions of J$^{-1}$s$^{-1}$, so reciprocal Joules per second.
Because $y$ describes the deformation of the director field relative to the length of a tactoid this would arguably require the collective reorientation of the colloidal particles in the nematic phase. Hence, we would expect $\tilde{\Gamma}_y$ to be inversely proportional to a rotational viscosity of the nematic phase. Relaxation of the aspect ratio of the droplets requires the transport of material in both the isotropic and nematic phases. It would therefore seem sensible to presume the relaxation rate $\tilde{\Gamma}_\alpha$ to be some average of the relevant viscosities of the two phases~\cite{doi1988}.  

For dimensional reasons this then implies that both rates should also be inversely proportional to some volume scale. Considering that the volume scale must represent the volume in which the viscous dissipation takes place, we conclude that both rates must be inversely proportional to the volume $V$ of a tactoid. Hence, we write for our dynamical equations in terms of the dimensionless free energy $f$ and tactoid volume $v$ as
\begin{equation}
\begin{aligned}
& \dfrac{\partial \alpha}{\partial t} = - \dfrac{\Gamma_{\alpha}}{v} \dfrac{\partial f}{\partial \alpha},
\\
&\dfrac{\partial y}{\partial t} = - \dfrac{\Gamma_{y}}{v} \dfrac{\partial f}{\partial y},
\end{aligned}
\label{scaledkineticequations}
\end{equation}
where $\Gamma_\alpha = \tilde{\Gamma}_\alpha V \sigma^2/(K_{11}-K_{24})$ and $\Gamma_y = \tilde{\Gamma}_y V \sigma^2/(K_{11}-K_{24})$ are our scaled fundamental rates with dimensions of reciprocal seconds that do no longer depend on the volume of a drop. Our approach should be equivalent to but extends that of Weirich \textit{et al.}~\cite{weirich2017liquid} and of Almohammadi \textit{et al.}~\cite{almohammadi2022shape}, who balance the rate of change in mechanical energy with the rate of energy dissipation for fixed values of our parameter $y$. 

To numerically integrate the two rate equations described in Eq.~\ref{scaledkineticequations}, we apply the (forward) Euler method function within the Mathematica package~\cite{wolfram1999mathematica}. 
As initial conditions, we use the solutions of $\partial f /\partial \alpha = \partial f / \partial y = 0$ in the absence of an external field, so for the case $\Sigma = 0$. A complete phase diagram describing the shape and director field of tactoids as a function of the dimensionless tactoid volume $v$ and the dimensionless electric field strength $\Sigma$ can be found in our earlier work~\cite{safdari2021}. Notice that we can make time dimensionless by defining $\tau = t\Gamma_\alpha$. In scaled time, our dynamical theory therefore introduces a single additional dimensionless group, namely $\gamma \equiv \Gamma_y / \Gamma_\alpha$. In our numerical evaluation of the kinetic equations, we choose to make use of a time-adaptive approach in order to efficiently deal with the fast and slow processes that turn out to characterize the response of tactoids. This means that we dynamically adjust the time step, referred to as $\mathrm{d}\tau$, during each evaluation step. Specifically, we steadily increase $\mathrm{d}\tau$ by a factor of 1.01  and assess the difference between the new values of $\alpha$ and $y$ relative to the ones of the previous time step. If the difference is less than 0.001 for $\alpha (t+\mathrm{d}\tau)-\alpha (t) < 0.001\ \alpha (t)$ or $y (t+\mathrm{d}\tau)-y(t)<0.001 \ y (t)$, we continue with a new value of the time step $d\tau$. Conversely, if one or both of the differences exceeds the mentioned thresholds, we divided $\mathrm{d}\tau$ by a factor of 50 in each subsequent time step. This adaptive approach is particularly useful in the late stages of simulation when changes in $\alpha$ and $y$ become exceedingly small, allowing us to expedite the numerical evaluation of the kinetic equation.

Having presented the main ingredients of our theory, we next discuss our most salient findings and compare our results with the experimental data of Metselaar \textit{\textit{et al.}}~\cite{Metselaar2017}.

\section{Results}
To determine the conditions under which elongated tactoids can be observed, even if only transiently, it is important to consider the following. As is now well established, small nematic tactoids tend to have a uniform director field, while for sufficiently large ones the director field is (for all intents and purposes) bipolar~\cite{Prinsen2003,Prinsen2004a,Jamali2015}. Switching on an electric field only acts to reorient tactoids with a uniform director field but does not affect their elongation. Bipolar ones reorient and do not become more elongated if their volume is smaller than some critical value~\cite{Metselaar2017}. If sufficiently large, bipolar tactoids may elongate substantially under the action of the field but only if somehow the director field remains bipolar and does somehow not immediately respond to the electric field ~\cite{safdari2021,kuhnhold2022structure}. This can only happen if (i) the ratio of the two fundamental relaxation rates $\gamma = \Gamma_y / \Gamma_\alpha$ is sufficiently small and (ii) the volume of the drop is sufficiently large for the electric field to be able to deform it. Making use of the scaling theory of our previous work~\cite{safdari2021}, we deduce by balancing the interfacial and Coulomb free energies that the latter happens if $v \gg \Sigma^{-1/3}$. Note that, typically, the anchoring strength varies from about $1.5$ to $6$~\cite{Jamali2015}. For the tactoids to be bipolar, we must in addition insist that $v \gg \omega^{-5/2}$~\cite{Prinsen2003}. The maximum aspect ratio we expect to find, provided that these conditions are met, is $R/r \sim \Sigma^{3/7} v^{1/7} \gg 1$ at the level of the scaling theory, so ignoring any constants of proportionality. This shows that the field strength more strongly impacts upon the maximum aspect ratio than the volume does. Our numerical evaluation of the kinetic equations confirms this.

\begin{figure}[!h]
\begin{center}	\includegraphics[width=8cm]{ 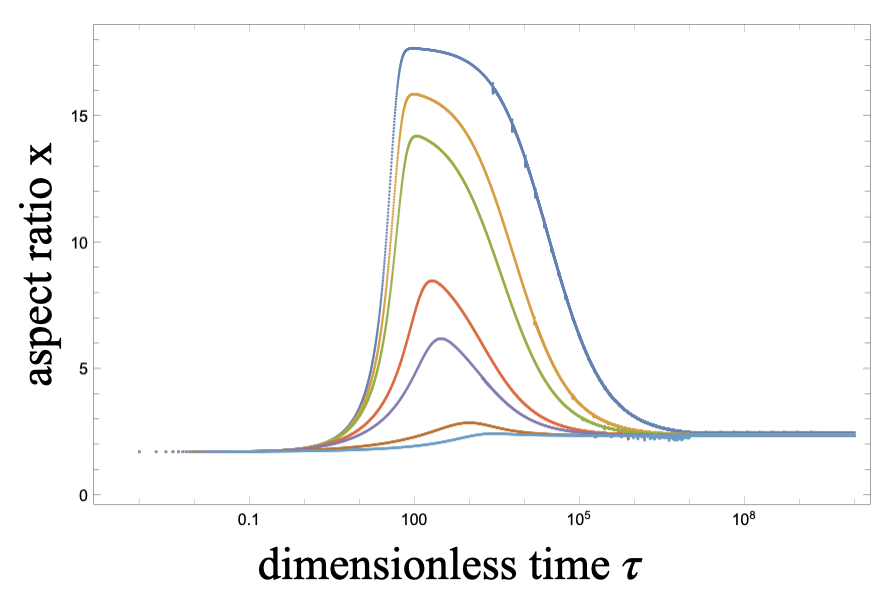}
\caption{Aspect ratio $x$ of a tactoid as a function of dimensionless time $\tau$ for different values of the ratio $\gamma$ of the fundamental relaxation rates associated with the response of the director field and that of the aspect ratio. From top to bottom: $\gamma=$ 0.01 (blue), 0.05 (yellow), 0.1 (green), 0.5 (red), 1 (purple), 10 (brown), 50 (light blue). The dimensionless volume of the droplet is $v = 10^5$, the anchoring strength $\omega=1.3$, the dimensionless strength of the electric field $\Sigma=65$, and the ratio of the bend and splay elastic constants $\kappa = 20$.}	
\centering 
\label{fig1-aspect-ratio}
\end{center}
\end{figure}

\begin{figure}[!h]
\begin{center}\includegraphics[width=8cm]{ 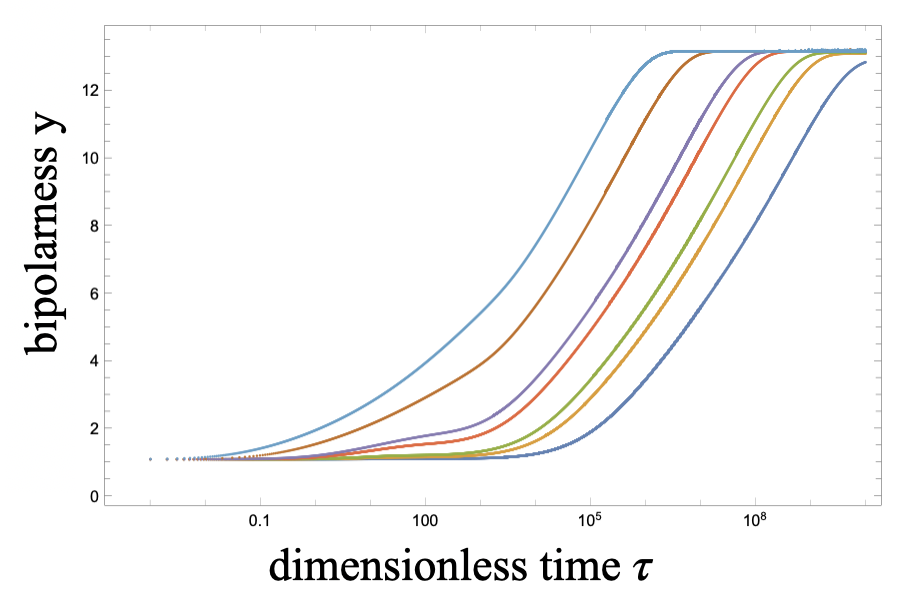}\caption{Bipolarness of a tactoid $y$ as a function of dimensionless time $\tau$ for the different values of the ratio $\gamma$ of the fundamental relaxation rates associated with the response of the director field and that of the aspect ratio. From bottom to top: $\gamma=$ 0.01 (blue), 0.05 (yellow), 0.1 (green), 0.5 (red), 1 (purple), 10 (brown), 50 (light blue). The dimensionless volume of the droplet is $v = 10^5$, the anchoring strength $\omega=1.3$, the dimensionless strength of the electric field $\Sigma=65$, and the ratio of the bend and splay elastic constants $\kappa = 20$.}	
    \centering 
    \label{fig1-bipolarness}
	\end{center}
\end{figure}

Now that we know under what conditions we might expect transients to arise, we first explore how the ratio of fundamental relaxation rates $\gamma$ influences the aspect ratio $x$ and bipolarness $y$ of tactoids.  Figure~\ref{fig1-aspect-ratio} shows the aspect ratio \textit{vs.} time for different values of $\gamma$, and fixed values of $v=10^5$, $\Sigma = 65$, $\omega = 1.3$, and $\kappa = 20$. The latter two values we obtain from the properties of chitin tactoids in the absence of an electric field~\cite{safdari2021}. For these values of the parameters, we have $v \Sigma^{3} \gg 1 $, and based on the scaling estimate we expect to see an overshoot of the aspect ratio of the order of 10 to 100 if $\gamma \ll 1$. Figure \ref{fig1-aspect-ratio} confirms this: the aspect ratio goes through a maximum before reaching its equilibrium value in the late stages of the process, and the maximum value reached increases as the magnitude of $\gamma$ decreases. Also, the smaller the value of $\gamma$, the longer lived the high-aspect-ratio states are. For late times, when the tactoids approach the true equilibrium state, the value of aspect ratio is not so large and does not depend on the ratio of $\gamma$. Hence, this confirms that large aspect ratios are possible only for smaller values of $\gamma$ when the director field has not yet fully equilibrated. Fig.~\ref{fig1-bipolarness}, showing the degree of bipolarness $y$ as a function of time, confirms this.

Fig.~\ref{fig1-bipolarness} shows that, not entirely unexpectedly, the smaller the value of $\gamma$ the more slowly the changes in bipolarness occur. This prevents the immediate alignment of particles and therefore that of the director field with the applied external electric field. The particles remain essentially enslaved to the surface anchoring until the internal dynamics allow them to break free from this and relax to the equilibrium value. Fig.~\ref{fig1-bipolarness} shows that there is some feedback between the two: whilst the bipolarness does increase monotonically with time, it does seem to develop a weak ``shoulder'' around the region where the aspect ratio of the droplets reaches its maximum. 
As time passes, the bipolarness of the tactoid increases reaching its equilibrium value, which leads to a reduction in the aspect ratio to a value consistent with our previous equilibrium results~ \cite{safdari2021}. This is the key reason for the presence of a maximum value for the aspect ratio as a function of time.

\begin{figure}[!h]
	\begin{center}\includegraphics[width=8.8cm]{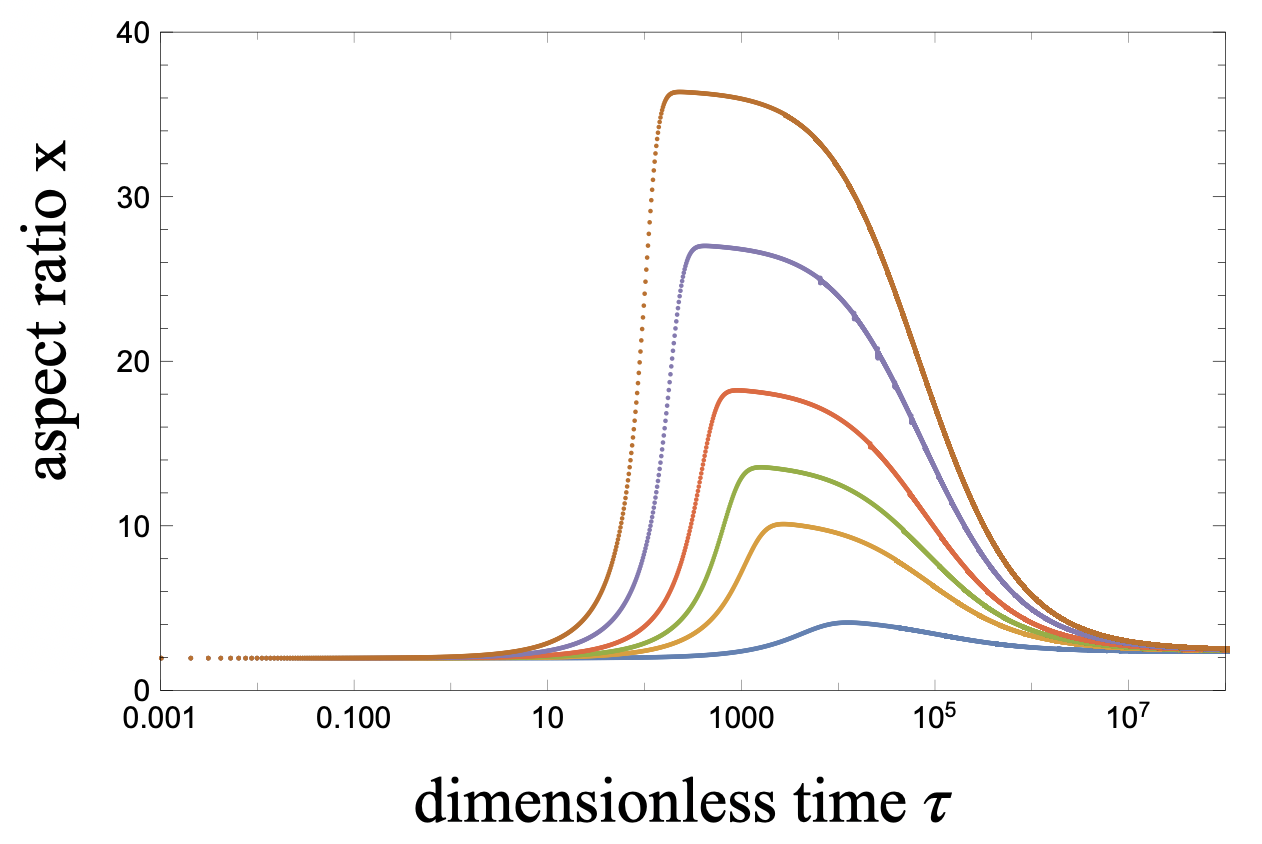}
    \caption{Aspect ratio $x$ of tactoid as a function of dimensionless time $\tau$ for different values of the dimensionless electric field strength $\Sigma=10,100,200,400,1000,2000$ (bottom to top). The dimensionless volume of the droplet is $v = 10^2$, the anchoring strength $\omega=1.3$, and the ratio of the bend and splay elastic constants $\kappa = 20$. The ratio of the fundamental relaxation rates was set at a value of $\gamma=1/15$. }	\centering 
	\label{ap-diff-electric-field}
	\end{center}
\end{figure}

We next investigate the impact of the strength of the external electric field on the shape and director-field structure of nematic tactoids. Figure \ref{ap-diff-electric-field} shows that the final, equilibrium value of the aspect ratio does not appreciably depend on the electric field strength, at least not for the values of $\Sigma=10, 100, 200, 400, 1000$ and $2000$ shown in the graph and the dimensionless volume of the drop $v=10^2$. This is consistent with previous equilibrium studies, showing that the final state is characterized by a more or less uniform director field and an aspect ratio that approaches the value of $2\sqrt{\omega}$ if the field is sufficiently strong~\cite{safdari2021,kuhnhold2022structure}. 
Still, transients with a large aspect ratio do arise with a maximum that increases with the strength of the electric field. We notice that the transients are very long-lived and more so the larger the field strength.
A careful review of the plots of the maximum value of aspect ratio found in Fig.~\ref{ap-diff-electric-field}) shows that it is proportional to $\Sigma^{0.42}$. The exponent is close to the prediction of $3/7$ mentioned earlier. Fig.~\ref{ap-diff-electric-field} also reveals that the time for a tactoid to reach its maximum elongation becomes shorter as the strength of the electric field becomes stronger. In fact, we find numerically that this time scales as $\Sigma^{-0.77}$.

\begin{figure}[!h]
\begin{center}\includegraphics[width=8.8cm]{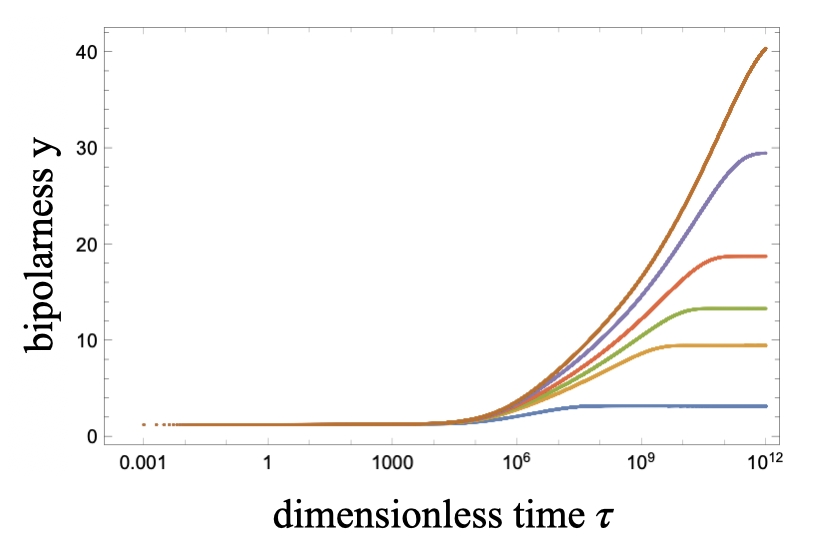}\caption{Bipolarness of tactoid $y$ as a function of the dimensionless time $\tau$ for the different electric field strength $\Sigma=10,100,200,400,1000,2000$ (bottom to top). The dimensionless volume of the droplet is $v = 10^2$, the anchoring strength $\omega=1.3$, and the ratio of the bend and splay elastic constants $\kappa = 20$. The ratio of the fundamental relaxation rates was set at a value of $\gamma=1/15$.}	\centering\label{bp-diff-electric-field}
	\end{center}
\end{figure}

 Let us now explore how the bipolarness $y$ of tactoids depends on the strength of the electric field $\Sigma$. As shown in Fig.~\ref{bp-diff-electric-field}, the equilibrium value of bipolarness increases with time, and more so the stronger the electric field. This means that the virtual point defects (the focal points of the extrapolated bipolar director field) move away from the poles of the tactoids and the director field becomes increasingly more homogeneous. A careful examination of our numerical results for late times shows that it increases as $\Sigma^{0.48}$. This is consistent with the equilibrium theory of Safdari \textit{\textit{et al.}}~\cite{safdari2021}, which predicts that the bipolarness grows with the field strength as $\Sigma^{0.5}$. Comparing Figs.~\ref{ap-diff-electric-field} and \ref{bp-diff-electric-field} also shows that the relaxation of the bipolarness to its equilibrium value is very much more sluggish than that of the aspect ratio. In fact, it is more sluggish than what we would expect based on the value of $\gamma$, which in the figure is equal to $1/15$. This is not all that surprising given that any director field with bipolarness above a value of, say, three is difficult to distinguish from a uniform director field. This implies that any response of the aspect ratio must be very small beyond that.

Finally, we investigate the effect of the volume of a tactoid on both its bipolarness and aspect ratio for a given external field and a given asymmetry in the relaxation dynamics of the director field and the droplet shape. From the earlier-mentioned scaling theory we expect that the director field is essentially uniform and the drop does not respond to any alignment field other than aligning along the field direction if the dimensionless volume $v$ is smaller than $\omega^{-5/2}$. For our choice of $\omega =1.3$, $v$ needs to be much larger than unity for it to have a noticeable degree of curvature of the director field. For the external field to be able to straighten out the curved director field, $v$ must be larger than about $\Sigma^{-5/4}$ according to the scaling theory. If we set $\Sigma=65$, this implies that a tactoid responds to that field as long the tactoid is bipolar, so $v\gg 1$. The maximum value of the aspect ratio scales then with the tactoid volume as $v^{1/7}$, which predicts a very weak dependence on the volume. 

\begin{figure}[!h]
\begin{center}\includegraphics[width=8.8cm]{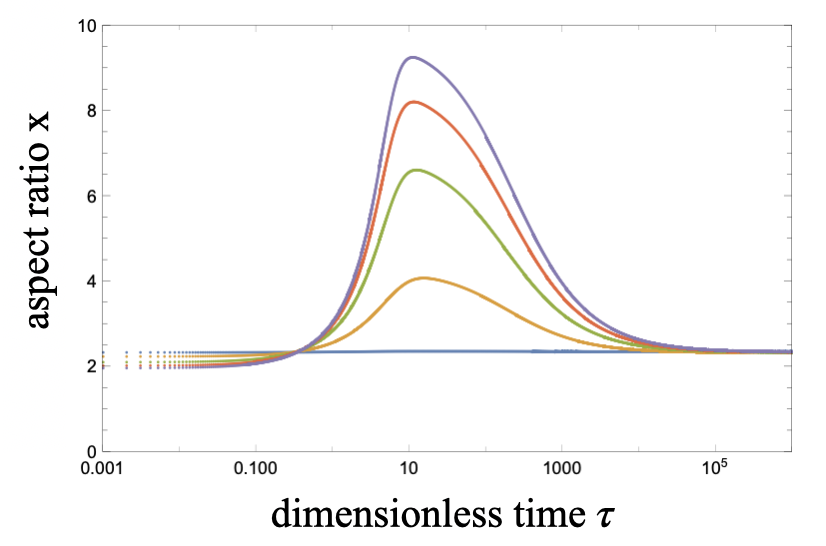}
 \caption{Aspect ratio of tactoids $x$ as a function of time for the different dimensionless volume $v=1,310,2340,5780,10000$ (bottom to top, central part). The electric field strength is $\Sigma = 65 $, the anchoring strength $\omega=1.3$, the ratio of the fundamental relaxation times $\gamma=1/15$ and the ratio of the bend and splay elastic constants $\kappa = 20$ }\centering\label{ap-diff-volume}\end{center}
\end{figure}

\begin{figure}[!h]
\begin{center}\includegraphics[width=8.8cm]    {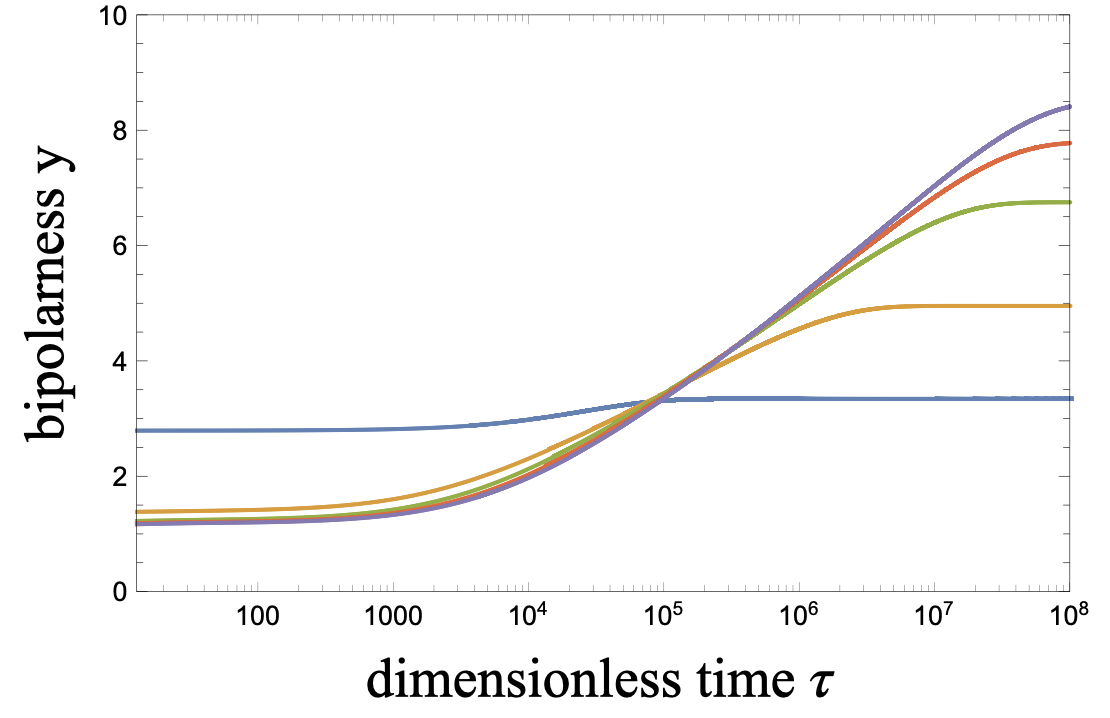}
 \caption{Bipolarness of tactoid $y$ as a function of the dimensionless time $\tau$ for different dimensionless volume  $v=1,310,2340,5780,10000$ (bottom to top on the right).  The electric field strength is $\Sigma = 65 $, the anchoring strength $\omega=1.3$, the ratio of the fundamental relaxation times $\gamma=1/15$ and the ratio of the bend and splay elastic constants $\kappa = 20$.}	
	\centering \label{bp-diff-volume}
	\end{center}
\end{figure}

Figures~\ref{ap-diff-volume} and~\ref{bp-diff-volume} show how a variation of the dimensionless volume for $v=1, 310, 2340$ and $5780$ affect the time-evolution of the aspect ratio and bipolarness following the switching on of the electric field for $\Sigma = 65$, $\omega=1.3$, $\gamma = 0.1$ and $\kappa = 20$.
The figure confirms once more that the equilibrium value of the aspect ratio of a tactoid does not, as expected, strongly depend on its volume~\cite{safdari2021,kuhnhold2022structure}. For $v=1$ the director field is almost uniform with $y\approx 3$, so for this volume, the external field has very little impact on both the aspect ratio and bipolarness. The larger tactoids are all essentially bipolar at time zero, with a bipolarness $y$ close to unity, and these do respond to the switching on of the field. The aspect ratio of this larger droplet does exhibit transient overshoots, while their bipolarness increases monotonically with time as the field straightens the director field. 

Figure~\ref{ap-diff-volume} confirms that the maximum value of the aspect ratio increases only weakly with the dimensionless volume that we varied over four decades in magnitude. The timescale required to reach this maximum value decreases with the volume of the drop but also not very strongly. The same can be said about the relaxation of the director field. The equilibrium value of the bipolarness does depend on the volume albeit not very sensitivly. According to the scaling theory of Safdari \textit{\textit{et al.}}, which is confirmed by numerical minimization of the free energy~\cite{safdari2021}, we expect $y$ to scale as $v^{1/6}$. The late-stage results shown in Fig.~\ref{bp-diff-volume} agree with this prediction, as they should.

After having investigated the response of the switching on of an electric field, we ask ourselves the question what happens when the electric field is turned off and the external field is removed? Specifically, we are interested in understanding the dynamics of the change in aspect ratio during the transition. Does it revert to its initial state swiftly? Or does it remain relatively unchanged? This is a relevant question, because  Metselaar \textit{\textit{et al.}} find that switching off of the field prior to full relaxation in the presence of the field the largest tactoids seem not to revert to the field-free aspect ratios in roughly the same amount time as when the field was switched on~\cite{Metselaar2017}. See also Figure~\ref{Experimental_of_Metselaar}, noting that quantitative data are not available.

\begin{figure}[!h]
\includegraphics[width=8cm]{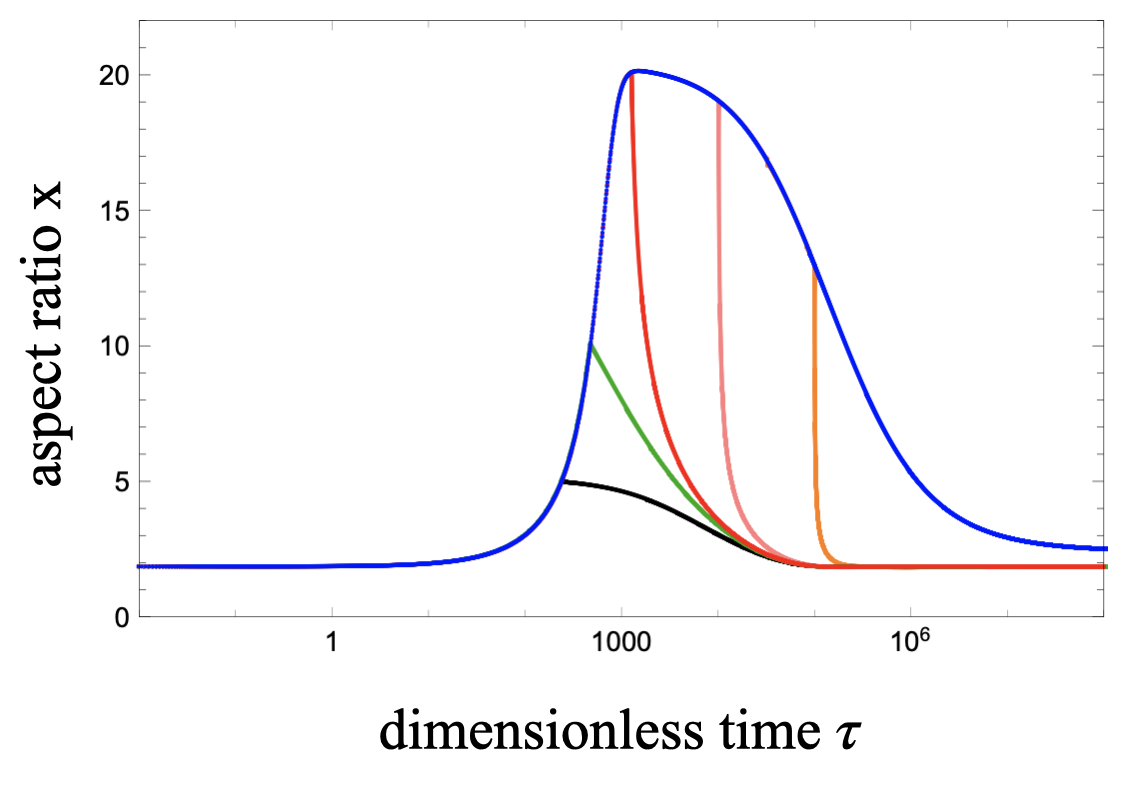}
 \caption{The evolution of the aspect ratio $x$ as a function of the dimensionless time $\tau$. The blue curve shows the response to switching on of the field at time zero. The curves in black, green, red, pink and orange show what happens after shutting of the field. The dimensionless volume of the droplet is $v = 5\cdot 10^3$, the anchoring strength $\omega=1.3$, the strength of the electric field is $\Sigma = 425$, the ratio of the bend and splay elastic constants $\kappa = 20$, and the ratio of the fundamental relaxation rates $\gamma=1/15$.  }	
\centering 
\label{On_vs_Off}
\end{figure}

To address these questions, we turn on and next turn off the electric field once the tactoids have elongated to a certain fraction of the maximum value for $\Sigma = 425$ and $v=5\cdot 10^3$. The results of our numerical calculations are presented in Fig.~\ref{On_vs_Off}. Consistent with the experiments of Metselaar {\it \textit{et al.}}~\cite{Metselaar2017}, we find that the tactoids take more time to approach their equilibrium configurations than the time required to elongate them. For the case shown, the difference in time amounts to  three or four orders of magnitude. We notice that the state of elongation at which the field is switched off does not seem to strongly influence the (dimensionless) relaxation time, which for all cases shown in the figure is about 70,000. The blue curve in Fig.~\ref{On_vs_Off} shows the situation when the electric field remains turned on, showing a considerably slower convergence to the equilibrium state in the presence of the field compared to the cases where the field was switched off, reverting to the equilibrium state in the absence of an alignment field. Both are much larger, though, than the time it takes to reach peak elongation. This illustrates that there are many very divergent timescales involved in the relaxation of tactoids, highlighting the non-linear character of the kinetics at hand.

In the next section, we apply the theory developed here in order to interpret the experimental results of Metselaar \textit{\textit{et al.}}~\cite{Metselaar2017}.

\section{Comparison with Experiment}
In a series of measurements, Metselaar {\it et al}.~recently measured the dimensions of tactoids in different batches of an aqueous dispersion of chitin nanocrystals as a function of time, following the application of a high-frequency AC electric  field~\cite{Metselaar2017}. Experiments were done for root-mean-square field strengths in the range from 160 V/mm to 450 V/mm and frequencies ranging from 300 to 700 kHz. Even though the findings between different experiments differed quantitatively, qualitatively the results are consistent with each other. 
Despite the small difference in the dielectric properties of the coexisting isotropic and nematic phases, which are dominated by the contribution from the aqueous solvent, it turns out that electric fields can significantly elongate tactoids at least if these are of sufficiently large volume. See also Figure~\ref{Experimental_of_Metselaar}. In our previous work~\cite{safdari2021}, focusing on the thermodynamic properties of tactoids in an electric field, we showed we could only get reasonable agreement between theory and experiment for the maximum elongation by invoking a restricted equilibrium. In this restricted equilibrium, we fixed the bipolarness of the tactoids to the equilibrium values obtained in the absence of the field.

Our aim here is to relax the restricted equilibrium, and adjust and calibrate our relaxational model in order to achieve the closest possible alignment with the experimental data. Given that the available data from different experiments agree only qualitatively, we seek to reproduce trends rather than achieve quantitative agreement. 
For this, we use model parameters obtained by fitting the theory to the available experimental data on field-free aspect ratio for a range of tactoid volumes varying four orders of magnitude~\cite{safdari2021}. In our previous work, we found the values for the anchoring strength $\omega = 1.3$, the ratio of bend to splay elastic constants $\kappa = 20$, and the extrapolation length $\left(K_{11}-K_{24}\right)/\sigma = 4$ $\mu$m. Hence, in the conversion of dimensionless volume $v$ to dimension-bearing volume $V$ we multiply $v$ by $64$ to change the unit of our dimensionless volumes to $\mu m^3$.


\begin{figure}[!h]
\includegraphics[width=8.8cm]{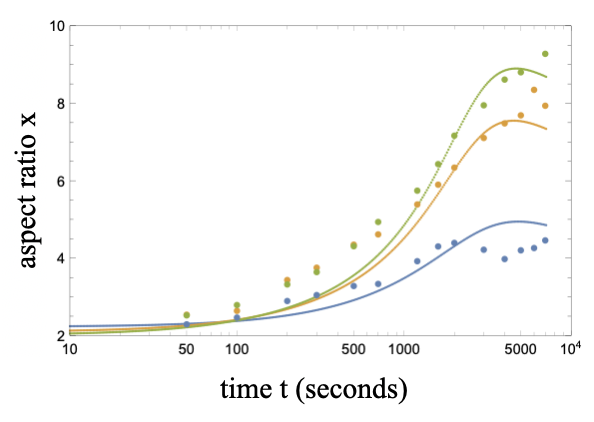}\caption{Time evolution of the aspect ratio of three droplets of different dimensionless volumes $v$, indicated by different colors: green for $v=5780$, orange for $v=2340$ and blue for $v=310$. Dots are experimental data points extracted from Ref.~\cite{Metselaar2017} and the drawn curves display our curve fits. We set the anchoring strength $\omega=1.3$, the ratio of bend to splay $\kappa=20$, the coefficient of the electric field strength $\Sigma=190$ the ratio of relxation rate $\gamma=0.5$ and $\Gamma_\alpha=1.6\times10^{-4}s^{-1}$. The $r^2$ values for the quality of the curve fits are 0.61, 0.91, and 0.94 for blue, orange, and green respectively.}	\label{dynamic_experiment_aspect ratio}
\end{figure}

Fig.~\ref{dynamic_experiment_aspect ratio} shows the aspect ratio $x=R/r$ \textit{vs.}~the actual time $t$ in seconds for three different tactoid volumes of $V\simeq 20\times 10^3$, $150 \times 10^3$ and $370\times 10^3$ $\mu$m$^3$.  
Shown are the ratios of lengths and widths of the chitin tactoids obtained experimentally by Metselaar {\it \textit{et al.}}~\cite{Metselaar2017}, together with the solid curves that are the result of our curve-fitted numerical solutions to the kinetic equations. Each color corresponds to a different volume. To find the best fit, we applied a hyperparameter grid search (on the parameters $\Sigma$, $\gamma$ and $\Gamma_\alpha$)~\cite{lerman1980fitting} and minimize the so-called cost function, $\sum (x_\mathrm{the}-x_\mathrm{exp})^2/N$ with N the number of data points, a measure for the mean-square distance between the aspect ratio of the theoretical prediction $x_\mathrm{the}$ and the experimental data points $x_\mathrm{exp}$. The values for the various parameters that we find are $\gamma=0.5$, $\Sigma=190$ and $\Gamma_\alpha=1.6\times10^{-4} s^{-1}$. 

We get reasonable agreement between theory and experiments, noting that we use the same values of all of the parameters except the volume of the drops, with coefficients of determination $r^2$ between $0.61$ and $0.95$. The curves clearly show that whilst the initial response time goes down with increasing droplet volume, final equilibration actually slows down and takes longer the larger the tactoids. Our curve-fitting procedure produces maxima that are not quite observed yet in the data, except perhaps for the smallest volume. It is important to point out that setting $\gamma \ll 1$ produces curves for which the maximum moves to much larger times. In fact, setting $\gamma =0$ and suppressing the maximum entirely produces curve fits that have much smaller values of $r^2$. In fact, a simple exponential relaxation, put forward in a slightly different context in Refs.~\cite{weirich2017liquid,almohammadi2022shape}, cannot describe the experiments in any satisfactory way (results not shown). We expect that for times much larger than, say, 7000 s, the experiments should show a downturn in the aspect ratio.


\begin{figure}[!h]
\includegraphics[width=7.5cm]{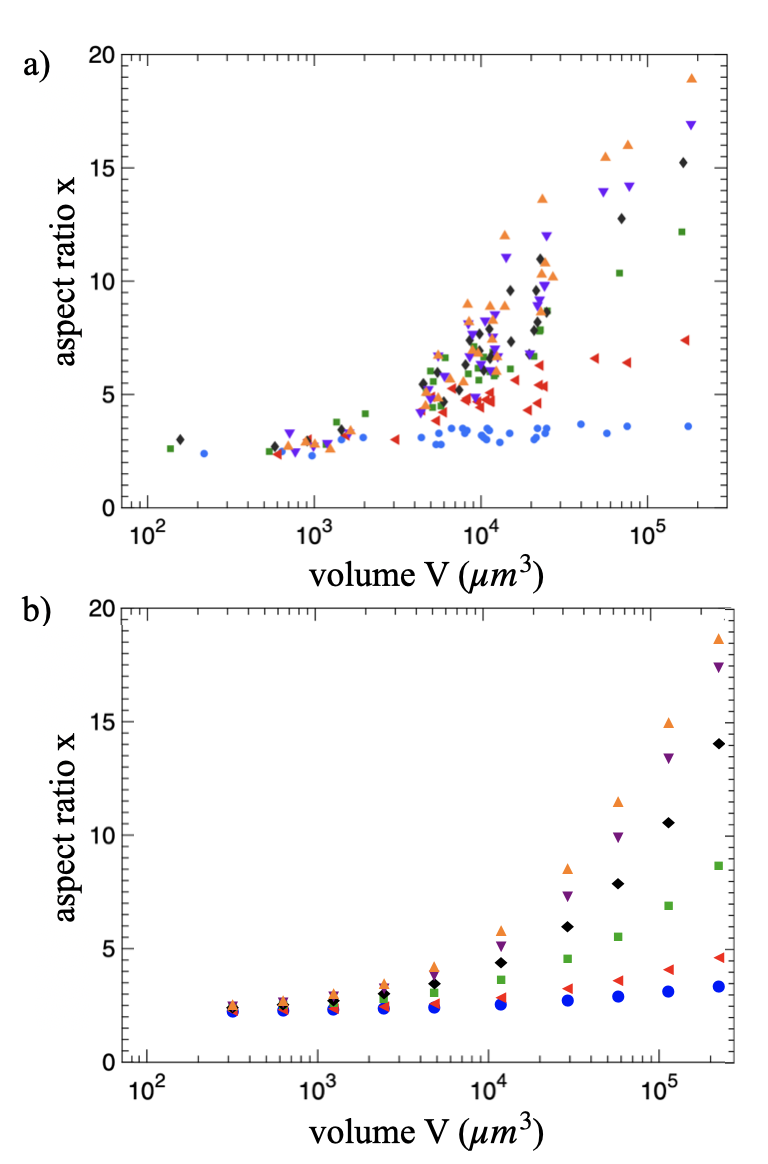}
\caption{Aspect ratio of tactoids $x$ as a function of their volume $V$ in $\mu$m$^3$ for different times. a) Experimental findings for chitin in water, taken from~\cite{Metselaar2017}. b) Our theoretical predictions are based on the curve fitting. Parameter values for the theoretical predictions: $\omega =1.3$, $\kappa=20$, $\Sigma=500$, $\gamma=1/20$ and $\Gamma_\alpha=1.6\times10^{-4}s^{-1}$. Results for the same moment in time are presented with the same color.  Blue circles: 120 seconds, red triangles: 210 s, green squares: 410 s, black diamonds: 640 s, purple triangles: 850 s,  orange triangles: 1080 s. See also the main text.} 
\centering 
\label{fig2}
\end{figure} 

In Figure~\ref{fig2}a) and b) we compare for the time evolution of tactoids ranging three orders in magnitude in size following the switching on of the electric field. Figure~\ref{fig2}a) shows the experimental findings for the aspect ratio $x$ of the tactoids as a function volume, and Figure~\ref{fig2}b) the results of our numerical simulations, where we used the value of $\Sigma = 500$ estimated from the restricted model of Ref.~\cite{safdari2021}. The other parameters we set at $\gamma = 0.05$ and $\Gamma_\alpha = 1.6 \times 10^{-4}$ s$^{-1}$ to obtain reasonable agreement (by eye). Note that the data set of the results of this figure is different from the one shown in Fig.~\ref{dynamic_experiment_aspect ratio}.
Different colors and symbols are used to denote various points in time. Agreement is quantitative. The figures show that for larger droplets, a gradual increase in elongation occurs over time due to the coupling to the external field. In contrast, smaller droplets exhibit minimal changes in their aspect ratios, because the Coulomb energy is not strong enough to be able to affect any changes in the droplet shape as it is strongly volume-dependent. The behavior predicted from our model closely mirrors the trends observed in the experimental data points, indicating an agreement between our simulations and the experimental observations.

\section{Conclusion}
In this study, we investigate theoretically the behavior of spindle-shaped nematic tactoid droplets in response to the switching on of an external electric field. For this, we set up a relaxational kinetic model based on a free energy landscape in terms of two reaction coordinates, extending our earlier work on the equilibrium structure of tactoids in an electric (or a magnetic) field~\cite{safdari2021}. These reaction coordinates are the aspect ratio and the degree of curvature of the director field of the droplet, which is presumed to be quasi bipolar and continuously interpolates between a uniform and a bipolar director field. The free energy is determined by a combination of the usual Frank-Oseen elasticity, a surface contribution that includes a preference for planar anchoring to the interface, and the coupling to the field via the dielectric anisotropy of the nematic phase.

We find that the elongation of the tactoids is purely a kinetic effect.  Indeed, it is not uncommon in nature for kinetic effects to transiently modify the shape and symmetry of a structure before it attains its final equilibrium form \cite{Panahandeh2022}. In this paper, we showed that the elongation of the tactoids is a transient effect and that they are only able to elongate strongly under the action of the electric field if the relaxation time of the degree of bipolarness of the director field is large compared to the relaxation time of the aspect ratio. Hence, the elongation of tactoids is only transient albeit that it may be very long-lived. Eventually, the aspect ratio returns to a much smaller equilibrium value predicted by equilibrium theory. Since equilibrium theory cannot explain strongly elongated configurations of tactoids observed in the experiments of Metselaar en collaborators~\cite{Metselaar2017}, we propose that it must be the approach to the transient state that the measurements have probed. This means that significantly longer measurement times are required to observe the return to equilibrium.

This conclusion is not as far-fetched as it may seem, because the aspect ratios of the largest drops observed in the experiments seem not yet to have reached the largest value at the latest times measured, as can in fact be concluded from the data shown in Figs. 10 and 11. 
That the relaxational dynamics of tactoids may exhibit a very long-time tail can also be deduced from the large scatter in aspect ratio found experimentally in a host of different systems~\cite{Jamali2015,Metselaar2017,puech2010,oakes2007,REVOL1992,kitzerow1994,park2014,wang2016,bagnani2018amyloid}. Perhaps even more convincingly, a recent study shows that the producing tactoids by means of nucleation-and-growth or making use of a microfluidic device result in different aspect ratios that seem stable for a very long period of time~\cite{almohammadi2023disentangling}. 

Our simple relaxational model captures the most salient features of the experimental findings. One of these is the observation that relatively small tactoids do not appreciably respond to an applied field, even if they are bipolar. The reason for this is that anchoring, Frank elastic and Coulomb free energies all scale with different powers of the volume of a tactoid~\cite{safdari2021}. We find that the larger the volume or the field strength, the shorter it takes for the aspect ratio to reach its largest value, but the longer it takes to relax back to its equilibrium value. 
Another significant aspect of this investigation is our finding that if the electric field is suddenly turned off, tactoids revert to their initial aspect ratio on a considerably shorter time scale than needed for the full equilibration to take place in response to switching on of the field. This seems also to agree with the experimental findings shown in Fig.~\ref{fig0}, although no quantitative data are available as far as we are aware.

It seems that the underlying fundamental relaxation time of the bipolarness of the director field of tactoids is much lower than that of the aspect ratio of the tactoids. This makes their aspect ratio become in a sense enslaved by the anchoring of the director field to the interface with the isotropic host suspension~\cite{Metselaar2017,safdari2021}. The reason for this behavior remains unclear. Within our model, it might reflect differences in the values of the viscosities that describe the relaxations of the shape and director field of a tactoid. These include five Leslie coefficients and the viscosity of the isotropic phase~\cite{de1993}. On the other hand, it may also point at the existence of complex interfacial relaxations not captured by current models. The issue in our view merits further study.

\acknowledgments{We thank Patrick Davidson and Ivan Drozov (Universit\'{e} Paris-Saclay) for helpful discussions and sharing of experimental data. M.S. and R.Z. acknowledge support from NSF DMR-2131963 and the University of California Multicampus Research Programs and Initiatives (Grant No. M21PR3267).}

\newpage
\bibliography{refs}
\end{document}